%% file: ESTER_review3.tex
\documentclass{svmult}
\usepackage{natbib}
\usepackage{graphicx}
\usepackage{MR_eas}
\usepackage{amsmath}

\usepackage{morefloats}
\usepackage[show]{sssrced}

\begin{document}

\title*{Ab initio modelling of steady rotating stars}
\author{Michel Rieutord and Francisco Espinosa Lara}
\date{\today}

\institute{Michel Rieutord \at Universit\'e de Toulouse; UPS-OMP; IRAP; Toulouse,
France and CNRS; IRAP; 14 avenue E. Belin, 31400 Toulouse, France
\email{Michel.Rieutord@irap.omp.eu}
\and Francisco Espinosa Lara \at Universit\'e de Toulouse; UPS-OMP; IRAP; Toulouse,
France and CNRS; IRAP; 14 avenue E. Belin, 31400 Toulouse, France
\email{Francisco.Espinosa@irap.omp.eu}}

\maketitle

\abstract{
Modelling isolated rotating stars at any rotation rate is a challenge for
the next generation of stellar models. These models will 
couple
dynamical aspects of rotating stars, like angular momentum and chemicals
transport,
with classical chemical evolution, gravitational contraction
or mass-loss. Such modelling needs to be achieved in two dimensions,
combining the calculation of the structure of the star, its mean flows
and the time-evolution of the whole.  We present here a first step
in this challenging programme. It leads to the first self-consistent
two-dimensional models of rotating stars in a steady state generated
by the ESTER code. In these models the structure (pressure, density
and temperature) and the flow fields are computed in a self-consistent
way allowing
the prediction of the 
differential rotation and the
associated meridian circulation of the stars. After a presentation
of the physical properties of such models and the numerical methods
at work, we give
the first grid of such models describing massive and
intermediate-mass stars for a selection of rotation rates up to 90\%
of the breakup angular velocity.
}

\section{Introduction}
\subsection{The astrophysical context}

In the last ten years rotation has become an essential part of stellar
models. It influences all stages of stellar evolution. For instance,
during  the formation process, a drastic reduction of the specific
angular momentum, by a factor of order $10^5$, is observed when matter
passes from the initial molecular cloud to the newly-born main-sequence
star. Later, during the evolution of the star on the main sequence,
many hydrodynamical instabilities, along with meridian circulations,
drive some mixing in the radiative zones. The effects of this mixing
are actually observed on the surface of many stars, which show elements
obviously synthesized in their core region. In addition, recent work by
\cite{FHT12} shows that rotation plays a role in the nucleosynthesis
of s-elements in massive stars.  Effects of rotation are also thought
to be important in understanding the evolution and “yields” of the
first generation of stars. Lacking in metals, these stars were more
compact than present stars and had also weaker winds. Thus, their
rotation was certainly faster than that of present day stars. It is
therefore crucial to master rotational effects in order to reconstruct
the history of metal enrichment in galaxies and to understand how the
observed stars have been influenced by the first generation.  Lastly,
we may mention an example of the importance of rotation in the end of
the life of stars: recent works on the gravitational collapse of massive
stars \cite[e.g.][]{Metzger_etal11} insist on the importance of the
combined effect of rotation and magnetic fields to model the explosion of
supernovae and the associated production of a gamma-ray burst.  The few
foregoing examples are just illustrative, since rotation influences
many other aspects of stellar physics like the oscillation spectrum,
mass-loss etc.

\subsection{The 1D-models}

Presently, rotation is included in spherically symmetric stellar
models through the approach proposed by \cite{zahn92}. Since these
models are one-dimensio\-nal all of the effects of rotation are averaged
on spheres. Thus differential rotation is only in the radial direction
(said to be shellular) and no meridian flow appears explicitly. Because
only the first terms of the spherical harmonic expansion of fields
are included, this modelling is valid at small rotation rates. An
equally important part of the models is that they assume the existence
of some small-scale turbulence in the radiative regions that efficiently
transport momentum horizontally (i.e. on isobars), erasing latitudinal
gradients of angular velocity.

Although limited to slow rotation and needing some {\it ad hoc} constants
(like turbulent diffusivities), this approach has the great merit of
allowing the use of 1D-models, which are now very precise as far as
microphysics is concerned. It thus allowed investigators to reproduce
many observed features of stars: surface abundance of lithium as a
function mass \cite[e.g.][]{CT99}, the (relative) high number of red
super-giants in low-metallicity galaxies \cite[e.g.][]{MM01}, or the
ratio of type~Ibc to type~II supernovae \cite[e.g.][]{MM05}, etc.

Although these examples reflect a truly successful modelling, the
understanding of the effects of rotation is still incomplete because in
many circumstances rotation is fast.  Stellar conditions thus do not
meet the requirements of the models and the use of Zahn's approach
becomes dubious.  Clearly, we are missing a more detailed view of
reality. For instance, no one knows the rotation rates that are allowed
in the foregoing 1D models.

\subsection{The history of 2D-models of rotating stars}

The solution to 
the deficiencies of 1D-models about rotation
will come from the use of multi-dimensional
models (two-dimensional at least), which include the mean-field
hydrodynamics along with the centrifugal distortion of the
star. Unfortunately, building such models turned out to be a very
difficult challenge. The story of this modelling is illuminating and
we briefly summarize it below, following the more detailed account
of \cite{R06c}.

The first step dates back to polytropic hydrostatic
models of \cite{james64}. An important step forward was
made by a US group around P.~Bodenheimer and J.~Ostriker
\cite[][]{OM68,OB68,Mark68,OH68,Jackson70,BO70,Boden71,Boden_Ost73}. Their
works are based on the Self-Consistent Field method which solves Poisson's
equation for the gravitational potential, $\Delta\phi = 4\pi G \rho$,
with the Green function, namely

\[ \phi(\vx) = -G\int
\frac{\rho(\vx')}{|\vx-\vx'|}\,d^3\vx' .  \]
This solution readily includes the boundary conditions at infinity for
the gravitational field.

Making general stellar models was impeded by many numerical
difficulties.  
The codes
could not deal with stellar masses less than 9~M$_\odot$
nor with very rapid rotation. An indication of the 
precision reached by
these models is given by the virial test (see below).  \cite{Jackson70}
found it to be around 4$\times$10$^{-3}$.

Soon after, M.~Clement took up the challenge and solved directly
the Poisson equation with a finite difference discretization
\cite[][]{clem74}. Results improved as testified by a better virial test
at 2$\times$10$^{-4}$.

Another series of work aimed at the construction of 2D-models
was published by the Japanese school led by Y.~Eriguchi
\cite[see][]{Eriguchi78,ES81}.  These works led to the first attempts to
account for the baroclinicity of radiative zones \cite[][]{UE94,UE95},
but the neglecting of viscosity imposed the prescription of some {\it
ad hoc} constraints. The series ended with the work of \cite{SHEM97}
who improved the microphysics but relaxed the dynamics, considering only
barotropic models.

To be complete, we need to mention the work of \cite{EM85,EM91} who
attacked the problem in a different way, using a mapping of the star. The
grid points are indeed mapped to the sphere through the relation

\[ 
r_i(\theta_k) = \zeta_i R_s(\theta_k)
\]
where $\zeta$ is a new radial variable spanning $[0,1]$. Finite
differences are used for both variables $\zeta$ and $\theta$.
The ensuing code was quite robust, being able to compute
configurations quite far from the pure sphere, but the precision
of calculation as monitored by the virial test remained around
$5\times10^{-4}$.

The more recent efforts on 2D modelling not related to the ESTER
project have been those of \cite{Roxburgh04}, motivated by the need
of models of rapidly rotating stars for asteroseismology, and those of
\cite{JMS04,JMS05} motivated by the discovery of the extreme flattening
of the Be star Achernar \cite[e.g.][]{DKJVONA05}. These two works use
barotropic models with an imposed rotation law. We note that \cite{JMS04}
used a new version of the Self-Consistent Field technique which leads to a
much more robust code, not restricted to a given mass range. These models
have also been used to investigate the acoustic oscillation spectrum of
rapidly rotating stars by \cite[][]{RTMJSM09,RMJSM09}. Lastly,
\cite{deupree11} worked out a series of 2D-models with uniform
rotation for masses between 1.625 and 8~M$_\odot$.

\section{The route to ideal models}

As may be observed, the weakest point of the foregoing models is
their lack of dynamics as well as of time-evolution. We recall that
any radiative region of a rotating star is pervaded by baroclinic flows
that appear in the first place as a differential rotation and a meridian
circulation \cite[e.g.][]{R06}. These flows control some turbulence there
and affect the whole star, playing an important role in the transport
of elements and momentum.

We may now wonder what would be the ideal model of an isolated rapidly
rotating star. This question was addressed in \cite{R06c} and we
reproduce here his discussion, which is still fully 
relevant:

``Such a model should describe the mean state of a star at any time of its
life and especially the new quantity specific to these stars: angular
momentum.

``Unlike a non-rotating star, which is a one-dimensional object (in
a large-scale description) which needs only scalar fields ([we] forget
magnetic fields), a rotating star is, at least, a two-dimensional object
with, at least, one vector field in addition to all scalar fields. Hence,
complexity increases not only by the multi-dimensional nature of the
model but also by the number of physical quantities to be determined. This
implies that the ideal model deals consistently with angular momentum and
especially the losses and gains through stellar winds and accretion. Such
a model should also take into account the baroclinicity of radiative
zones and there, the anisotropic turbulence which appears through shear
instabilities; it should also include a mean-field theory of convection
to forecast Reynolds stresses and heat flux. Of course, observers would
like to know the emissivity of the atmosphere as a function of latitude
(if they use interferometry) or as a function of wavelength if they do
spectroscopy. But if they do asteroseismology they surely wish to know
the eigenspectrum of these objects.

``The foregoing points show that progress in the understanding of rotating
stars needs also some advances in the following questions of stellar physics:

\begin{itemize}
\item How angular momentum is distributed in a star and how is it input
or output with what consequences ?

\item The immediately following question concerns the nature of the
Reynolds stresses in the convective and radiative zones.

\item Then, what is the baroclinic state of the radiative regions ?

\item Similarly, the atmosphere is in a  baroclinic state and cannot
be at rest: how strong are the differential rotation and the meridional
currents? Does the atmosphere develop strong azimuthal winds streaming
around the star like Jupiter's winds?

\item Gravity darkening can be so efficient that equatorial regions
are cool enough to develop convection; this raises the question of the
latitude dependence of emissivity of the atmosphere beyond the von~Zeipel
model \cite[see the attempt of][]{LDS06}.

\item \relax 
[we] did not mention magnetic fields. Clearly, they multiply the
number of problems and first steps should ignore them if possible."
\end{itemize}

\section{Setting the problem}

Many of the foregoing questions can be answered by
steady rotating stars,
which do not evolve, neither dynamically by losing mass
and angular momentum, nor by gravitationally contracting, nor by nuclear
evolution. As a first step towards the general models, we concentrate
on this simpler problem.

\subsection{Equations for a steady rotating star}

We consider a lonely rotating star in a steady state. The star is
governed by the following equations for macroscopic quantities:
\begin{equation}
  \label{basiceq}
  \left\{
  \begin{aligned}
    & \Delta\phi = 4\pi G\rho \\
    & \rho T \vv\cdot\na s = -\Div\vF + \eps_*\\
    & \rho (2\vO_*\wedge\vv + \vv\cdot\na\vv) = -\na P
    -\rho\na(\phi-\textstyle{\demi}\Omega_*^2s^2)+\vF_v\\
    & \Div(\rho\vv) = 0.
  \end{aligned}
  \right.
\end{equation}
The first equation is Poisson's equation which relates the mass
distribution ($\rho$ is the density) and the gravitational potential
$\phi$ ($G$ is the gravitation constant).  The second equation is the
heat equation, which relates the advection of entropy $s$ by the velocity
field $\vv$ when nuclear heat sources $\eps_*$ are present, with
the heat flux $\vF$. The third equation is the momentum equation written in
a frame rotating at angular velocity $\vO_*$. $P$ is the pressure field
and $\vF_v$ the viscous force.  Finally the last equation 
ensures mass conservation.

These equations need to be completed by those describing the microphysics
and the transport properties of stellar matter. We propose to describe the
heat flux with

\[ 
\vF = -\khi_r\na T -\frac{\khi_{\rm turb}T}{{\cal R}_M}\na s
\]
where $\khi_r$ is the radiative conductivity and $\khi_{\rm turb}$
a turbulent heat conductivity. In this expression the second term is
assumed to represent the convective heat flux, which is supposed to be
controlled by the entropy gradient. ${\cal R}_M={\cal R}/{\cal M}$ where
${\cal R}$ is the ideal gas constant and ${\cal M}$ the mean molecular
mass of the fluid.

A realistic modelling of the viscous force would be derived from the
turbulent Reynolds stresses, however before reaching this long term
goal we assume the simple model of a constant dynamical viscosity $\mu$,
namely:
\[ 
\vF_v = \mu(\Delta\vv + \frac{1}{3}\na\Div\vv).
\]

The microphysics is completed by the three expressions
\begin{equation}
  \left\{
  \begin{aligned}
    & P\equiv P(\rho,T)\\
    & \kappa \equiv \kappa(\rho,T)\\
    & \eps_* \equiv \eps_*(\rho,T)
  \end{aligned}
  \right.
\end{equation}
which respectively give the equation of state, the opacities and
the nuclear heat generation. We recall that in radiative diffusive
equilibrium, heat conductivity is related to opacity by
\[ 
\khi_r = \frac{16\sigma T^3}{3\kappa\rho},
\]
where $\sigma$ is the Stefan-Boltzmann constant.

\subsection{Boundary conditions --- Angular momentum condition}

The previous system of partial differential equations needs to be
completed by boundary conditions which affect the gravitational potential,
the velocity field, the pressure and temperature.

At the star centre these conditions are simply that all functions should
be regular. At the surface things are more difficult. First because the
surface of a star is ill-defined. To simplify, we shall assume that the
surface is an isobar or an equipotential. If the properties of the model
are independent of the chosen surface, then we may say that the choice
of the bounding surface is not crucial. On this surface we have to:
\begin{itemize}

\item match the gravitational potential to that in the vacuum, which
vanishes at infinity;

\item impose stress-free conditions on the velocity field, namely that
  \[ 
  \vv\cdot \vn =0 \quad {\rm and}\quad ([\sigma]\vn)\wedge\vn =\vzero
  \]
  where $[\sigma]$ is the stress tensor.
  The fluid is assumed to not flow outside the star, and to feel
  no horizontal stress ($\vn$ is the outer normal of the star);

\item specify the pressure on the surface. A simple choice is
  $P_s=\frac{2}{3}\frac{\overline{g}}{\overline{\kappa}}$, where
  $\overline{\kappa}$ and $\overline{g}$ are the averaged opacity and
  gravity on this surface;

\item impose the temperature boundary conditions. To simplify we assume that the star
  radiates locally as a black body. Therefore we ask
  \[ 
  \vn\cdot\na T + T/L_T=0
  \]
  where $L_T$ is specifying the scale of temperature variations near
  the surface. We note that if the diffusion approximation is used then
  $L_T=16/(3\rho\kappa)$, $\kappa$ being the opacity. 
\end{itemize}

We note that this problem, as formulated by Eqs.~\eq{basiceq},
is not fully constrained because the total angular momentum is not
specified. We have specified the rotation rate of the frame where the
solution is computed but this is not specifying the rotation rate of
matter, which is the combination of both $\vv$ and the solid body rotation
of the frame. To remove this degeneracy we may either specify the total
angular momentum of the matter or specify the equatorial velocity of
the star. In the first case we may write
\[ 
\intvol r\sth\rho u_\varphi \,dV = 0 
\]
and 
\[ 
v_\varphi(r=R,\theta=\pi/2) = 0
\]
in the second case.  The first condition just specifies that all the
angular momentum of the star is in the solid body rotation of the frame,
while the second just says that the frame is rotating at the equatorial
angular velocity.

\subsection{Scalings}

As formulated by \eq{basiceq}, the problem is that of the steady flow
of a self-gravitating compressible gas subject to some heat sources. As
all fluid mechanics problem we first need to choose the various relevant
scales that feature the physical quantities. These scales are used to
rewrite the set of equations with dimensionless variables. We converged 
to the following set:
\begin{center}
  \parbox{10cm}{
    \noindent
    Length scale: equatorial radius \dotfill $R$ \\
    Time scale \dotfill $(R^2/\RM T_c)^{1/2}$ \\
    Velocity scale \dotfill $V=\sqrt{\RM T_c}$ \\
    Density scale: central density \dotfill $\rho_c$ \\
    Temperature scale: central temperature \dotfill $T_c$ \\
    Pressure scale \dotfill $\RM\rho_c T_c$ \\
    Entropy scale \dotfill ${\cal R}_M$ \\
    Potential scale \dotfill  $4\pi G R^2\rho_c$
  }
\end{center}

This scaling uses the sound travel time as the time scale, and
as a consequence the velocity field is scaled by a central sound
velocity. Accordingly, the potential scale would be $V^2$, but we prefer
to use $4\pi G R^2\rho_c$, which comes from the free-fall time scale.

\subsection{Dimensionless equations and numbers}

Using the foregoing scaling leads to the following dimensionless equations
\begin{equation}
  \label{dimless_eq}
  \left\{ 
  \begin{aligned}
    & 
    \begin{split}
      \rho\lp 2\Omega \ez\wedge\vu + (\vu\cdot\na)\vu\rp 
      = 
      -\na p - \rho\na(\Lambda_p\psi - \demi\Omega^2s^2) \\
      {}+ E\lp\Delta\vu + \textstyle{\frac{1}{3}}\na\Div\vu\rp
    \end{split} \\
    &
    \rho T\vu\cdot\na s = \Div\lp \khi\na T+\khi_tT\na s\rp +\eps\\
    &
    \Div\rho\vu = 0\\
    &
    \Delta\psi = \rho\\
  \end{aligned}
  \right.
\end{equation}
with the numbers
\[ 
\Omega = \frac{\Omega_*R}{\sqrt{{\cal R}_M T_c}},\qquad E =
\frac{\mu}{\rho_c R\sqrt{{\cal R}_M T_c}}, \qquad \Lambda_p =
\frac{4\pi GR^2\rho_c}{{\cal R}_M T_c}
\]
and the dimensionless functions
\[
\psi = \frac{\phi_*}{4\pi G R^2\rho_c},\qquad
\khi = \frac{\khi_r(\rho,T)}{{\cal R}_M\rho_cR\sqrt{{\cal R}_M T_c}},
\qquad \eps =\frac{R\eps_*(\rho,T)}{\rho_c ({\cal R}_M T_c)^{3/2}}.
\]
Although the critical angular velocity can only be computed after the
completion of the calculation, it is useful to define the pseudo-critical
angular velocity $\Omega_c=\sqrt{4\pi G\rho_c}$,
\[ 
\frac{\Omega_*}{\Omega_c} = \frac{\Omega}{\sqrt{\Lambda_p}}
\]
and introduce other dimensionless functions, namely
\[ 
\epsb=\eps/\eps(0), \qquad \khib=\khi/\khi(0), \qquad
\Lambda_T = \frac{\eps(0)}{\khi(0)}=
\frac{R^2\eps_*(\rho_c,T_c)}{T_c\khi_r(\rho_c,T_c)}.
\]

\subsection{Global parameters, $\rho_c, T_c, R$ of the star model}

Let us suppose that we know the mass of the star $M$, how can the
foregoing equations be used to determine the stellar quantities,
especially $\rho_c, T_c$ and $R$?

If we have solved the problem, we observe that $\Lambda_p$ is known,
which gives a first relation between $\rho_c$, $T_c$ and $R^2$.
In the same manner, the central value of $\varepsilon$ gives a relation
between $\rho_c$ and $T_c$:
\[ 
\eps(0) =\frac{R \eps_*(\rho_c,T_c)}{\rho_c(\RM T_c)^{3/2}}.
\]

Thus, taking into account that the mass of the star expresses as a
function of $R$ and $\rho_c$, namely
\[ 
M=\rho_cR^3\intvol \rho \,dV,
\]
the expressions of $\Lambda_p$ and $\varepsilon(0)$ completely determine
the parameters of the stars, i.e. the radius, the central temperature
and central density.

\subsection{Numerical method}

The resolution of the set of equations (\ref{dimless_eq}) imposes
some numerical challenges.  First, the shape of a rotating star is
spheroidal and not known a priori. We thus use a mapping of
coordinates adapted to this geometry and modify it during
the calculation.  Secondly, as the problem is two-dimensional, its size
quickly grows with the resolution, thus 
imposing the use of an efficient numerical method that should be, at
the same time, precise enough to deal with velocity fields. Finally, as
the problem is non-linear, we solve it using an iterative 
procedure.

\subsubsection{Computational domain and mapping of coordinates}

Following \citet{BGM98}, we use a mapping of the coordinates
$r(\zeta,\theta)$ adapted to the geometry of the star. The new
spheroidal coordinates $(\zeta,\theta)$, are defined in
such a way that $\zeta=1$ represents the surface of the star, $\theta$
being the colatitude. Thus doing, the surface of the star is a surface
of coordinate, a property that very much simplifies the implementation
of boundary conditions.

The star can be subdivided in several domains, and
in each domain the relation between spherical and spheroidal coordinates
is given by the general expression
\begin{equation}
  r=a_i\zeta+A_i(\zeta)[R_{i+1}(\theta)-a_i\eta_{i+1}]+B_i(\zeta)[R_i(\theta)-a_i\eta_i]
  \quad \eta_i\leq\zeta\leq\eta_{i+1},
\end{equation}
where $B_i(\zeta)=1-A_i(\zeta)$. Here
$R_i(\theta)$ and $R_{i+1}(\theta)$
represent the internal and external boundary of the domain respectively,
and $a_i$, $A_i(\zeta)$ are chosen to satisfy some properties at the
boundaries between the different domains.  In particular, we want
$$
r(\zeta=\eta_i,\theta)=R_i(\theta) \quad {\rm and} \quad 
r(\zeta=\eta_{i+1},\theta)=R_{i+1}(\theta),
$$
then it should be that $A_i(\eta_i)=0$ and $A_i(\eta_{i+1})=1$. The value of
the parameters $a_i$ should be such that $r$ is monotonically increasing
with $\zeta$.

A nice property of this mapping is that it reduces to the spherical
coordinates near the centre. The use of a spherical harmonic
expansion of the unknowns is therefore possible, and the asymptotic
properties of the solutions near the centre are well-known. The
decomposition of the stellar domain into multi-subdomains improves the
efficiency of the spectral methods to be used, especially in dealing
with discontinuities (interface between a convective and radiative
region) and in dealing with the large pressure and density variations.

\begin{figure}[t]
\includegraphics[width=\textwidth]{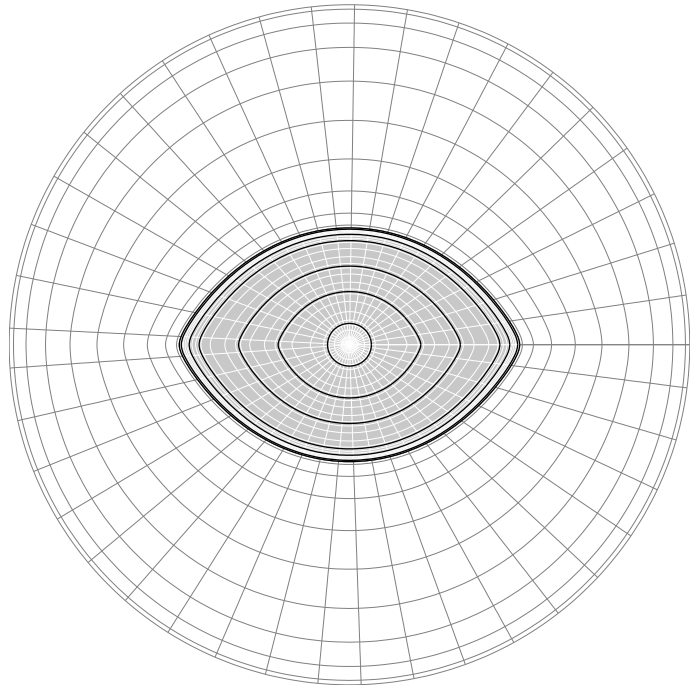}
\caption{The computational domain: the star at the centre is divided
into sub-domains (layers) that ease the computations. It is surrounded
by a vacuum domain limited by a sphere which allows an easy
implementation of the boundary conditions on the gravitational
potential.}
\label{fig_mapping}
\end{figure}

As we have already mentioned, at the stellar surface, the gravitational
potential must match the vacuum solution vanishing at infinity. There is
no easy way of imposing this condition on a surface with arbitrary shape.
To circumvent this difficulty, we follow \cite{BGM98} who use an
additional domain such that $1\leq \zeta\leq 2$ and whose outer boundary
is a sphere as shown in Fig.~\ref{fig_mapping}. On this sphere bounding the
computational domain, the gravitational potential can meet simple
boundary conditions for each of its spherical harmonics, namely
\[ 
\dr{\phi_\ell}+\frac{(\ell+1)\phi_\ell}{r}=0
\]
which ensure the matching with a field vanishing at infinity.
Fig.~\ref{fig_mapping} shows the full computational domain along with
the key surfaces of the mapping.

To write down the equations \eq{dimless_eq}, the operators should be
written with the spheroidal coordinates.  Let first us recall that
the relation between these coordinates and the classical spherical
coordinates is
\beq r\equiv r(\zeta,\theta'),
\quad \theta=\theta', \quad \varphi=\varphi'.
\eeqn{tcoordb}
In order to manipulate these new coordinates, and especially the
vectors, we use the natural basis defined by
\[
\vE_\zeta = \dzeta{\vr},\quad \vE_\theta=\dtheta{\vr}, \quad
\vE_\varphi=\dphi{\vr}.
\]
Making explicit
these definitions, we can express the covariant and
contravariant vectors, namely:
\[
\vE_\zeta = r_\zeta\er, \quad \vE_\theta=r_\theta\er+r\etheta, \quad
\vE_\varphi=r\sth\ephi
\]
\[
\vE^\zeta = \frac{\er}{r_\zeta}-\frac{r_\theta}{rr_\zeta}\etheta,
 \quad \vE^\theta=\frac{\etheta}{r}, \quad
\vE^\varphi=\frac{\ephi}{r\sth}
\]
as functions of the usual unit vectors associated with spherical
coordinates ($\er,\etheta,\ephi$); we set:
\[
r_\zeta = \dzeta{r}, \quad r_\theta=\dtheta{r}. 
\]
Passing from the spherical to the spheroidal coordinates may be done
using the following relations:
\beq \dr{f}=\frac{1}{r_\zeta}\dzeta{f}, \qquad
\dtheta{f}=\dthetap{f} - \frac{r_\theta}{r_\zeta}\dzeta{f}, \qquad
\dphi{f} = \dphip{f}\, .
\eeq

These expressions give the metric tensor whose covariant
($g_{ij}=\vE_i\cdot\vE_j$) and contravariant ($g^{ij}=\vE^i\cdot\vE^j$)
components are:
\beq
g_{\zeta\zeta}=r_{\zeta}^{2}, \qquad
g_{\zeta\theta}=r_{\zeta} r_{\theta},\qquad
g_{\theta\theta}=r^2+r_{\theta}^{2}, \qquad
g_{\varphi\varphi}=r^{2} \sin^2\!\theta
\eeq
\beq g^{\zeta\zeta}=\frac{r^2+r_\theta^2}{r^2r_\zeta^2}, \qquad
g^{\zeta\theta}=-\frac{r_\theta}{r^2r_\zeta}, \qquad
g^{\theta\theta}=\frac{1}{r^2}, \qquad
g^{\varphi\varphi}=\frac{1}{r^2\sin^2\theta}
\eeq
and
$g_{\zeta\varphi}=g_{\theta\varphi}=g^{\zeta\varphi}=g^{\theta\varphi}=0$.
At $r=0$, $g^{\theta\theta}, g^{\varphi\varphi}$ are
singular.  The metric determinant is such that
\[ 
\sqrt{|g|} = r^2r_\zeta\sth \andet |\eps^{ijk}|= \frac{1}{\sqrt{|g|}}
\]
The volume element is given by
\[ 
dV 
= \sqrt{|g|} \,d\zeta \,d\theta \,d\varphi 
= r^2r_\zeta\sth \,d\zeta \,d\theta \,d\varphi
= r^2r_\zeta  \,d\zeta \,d\Omega.
\]
As an example, the Poisson equation for the gravitational potential
takes the form
\begin{align*}
  \Delta\psi 
  &=
    g^{\zeta\zeta}\frac{\partial^2\psi}{\partial\zeta^2}+
    \left[\frac{2}{rr_\zeta}\left(1+\frac{r_\theta r_{\zeta\theta}}{rr_\zeta}\right)-
      g^{\zeta\zeta}\frac{r_{\zeta\zeta}}{r_\zeta}-
      \frac{r_{\theta\theta}+r_{\theta}\cot\theta}{r^2r_\zeta}\right]
    \frac{\partial\psi}{\partial\zeta}-\\
  & \qquad {} - \frac{2r_\theta}{r^2r_\zeta}
    \frac{\partial^2\psi}{\partial\zeta\partial\theta}+
    \frac{1}{r^2}\left[\frac{1}{\sin\theta}\frac{\partial}{\partial\theta}
      \left(\sin\theta\frac{\partial\psi}{\partial\theta}\right)+
    \frac{1}{\sin^2\theta}\frac{\partial^2\psi}{\partial\varphi^2}\right]\\
  &= \rho ,
\end{align*}
where $r_{\zeta\zeta}=\frac{\partial^2 r}{\partial\zeta^2}$,
 $r_{\theta\theta}=\frac{\partial^2 r}{\partial\theta^2}$ and
$r_{\zeta\theta}=\frac{\partial^2 r}{\partial\zeta\partial\theta}$.

\subsubsection{Numerical method}

As we have mentioned earlier, due to the 2D nature of the problem, the
size of the calculation will increase very quickly with the number of grid
points. This motivates the use of a high order method that can achieve
high precision with low resolution grids, optimizing computation
time and memory requirements.

An additional difficulty comes from the fact that the overall structure
of the star (profile of pressure, temperature, etc.) and its dynamics
(rotation, meridional circulation) must be computed simultaneously. This
requires a great precision and the ability to calculate higher order
derivatives of some variables.

Spectral methods are specially well-suited for this kind of 
problem
\cite[][]{Grandclement06,CHQZ06}. These methods expand the solutions on a basis
of orthogonal functions. The approximation of the solution using $n$
basis functions will be
\begin{equation}
  \phi^{(n)}(x)=\sum_{l=0}^{n-1} a_l P_l(x).
\end{equation}
The basis functions $P_l(x)$ 
used are usually a set of orthogonal polynomials, such as the 
Legendre or Chebyshev polynomials.
In our case it can be shown that
if $\phi(x)$ is infinitely differentiable, the approximate function
$\phi^{(n)}(x)$ will converge to the exact solution $\phi(x)$ faster
than any power of  the grid resolution $h$.
This expresses the fact that, in spectral methods,
the error decreases exponentially with the number of basis functions $n$
\cite[][]{fornberg}.
The $i$-th derivative of a function is approximated by
\begin{equation}
\left(\frac{\D^i\phi}{\D x^i}\right)^{(n)}
= \sum_{l=0}^{n-1} a_l \frac{\D^i P_l}{\D x^i}.
\end{equation}

Due to the non-linearity of the equations that we have to solve,
we have to calculate the product of two variables of the
model in an efficient way. Unfortunately, multiplication cannot be
easily performed using the spectral coefficients, as it involves a
convolution in the transformed space. This can be solved by using
an alternate approach of spectral methods, with the same properties,
called pseudospectral collocation methods. In a pseudospectral method
the variables are not represented by their spectral coefficients, but by
their values at certain points $x_i$ called the \emph{collocation points}.
Then, all the calculations can be performed in the real space.

The basis functions are orthogonal against some scalar product
\linebreak \mbox{$\langle P_l,P_m \rangle = \delta_{lm}$}.
Then, we can get the spectral coefficients
$a_l$ by 
$$
a_l = \langle \phi(x), P_l(x) \rangle .
$$ 
The scalar product (usually a
weighted integral over the interval) can be calculated using the gaussian
quadrature formula associated with the family of orthogonal polynomials
$P_l$ $$a_l=\sum_{j=0}^{n-1}w_jP_l(x_j)\phi(x_j),$$ where $x_j$
and $w_j$ are the nodes and weights of the gaussian quadrature. Note
that $x_j$ are the collocation points. Then, the first derivative at
the collocation points can be obtained as
\begin{align*}
\phi'(x_i) 
& = \sum_{l=0}^{n-1}\biggl(\sum_{j=0}^{n-1}w_jP_l(x_j)\phi(x_j)\biggr)P'_l(x_i)\\
& = \sum_{j=0}^{n-1}\biggl(\sum_{l=0}^{n-1}w_jP_l(x_j)P'_l(x_i)\biggr)\phi(x_j)\\
& = \sum_{j=0}^{n-1} D_{ij}\phi(x_j)
\end{align*}
where $D_{ij}$ is the differentiation matrix.

We use this procedure in the radial and latitudinal directions to
transform the original system of non-linear partial differential equations
in a system of non-linear algebraic equations.  For the latitudinal
direction, we use Legendre polynomials as the basis functions, while
for the radial  direction we use Chebyshev polynomials associated with
the Gauss-Lobatto collocation points. This grid includes the points at
the extrema in order to deal with boundary conditions.

The nice convergence properties of spectral and pseudospectral
methods are only valid for smooth functions which are infinitely
differentiable. However, it is known that inside a star there will be some
discontinuities, as for example, at the boundary between a convective
core and a radiative envelope. This difficulty is solved by using a
multi-domain approach, in a way that the variables are continuous and
differentiable within each domain but not necessarily at the boundaries
between different domains.

\subsubsection{Iterative procedure}

The system of algebraic equations resulting from the discretization of
the problem is non-linear and is solved using an iterative method.
For that we either use the well-known Newton's method or a relaxation
method. For Newton's method we write the problem
in the form
\begin{equation}
\vec F(\vec x)=\vec 0 ,
\end{equation}
where the vector function $\vec F$ represents the equations that we
want to solve and $\vec x$ is the vector containing all the independent
variables of the problem (pressure, temperature, \ldots) including the
shape of the surface which is not known a priori.  The equations are
linearized using the Jacobian matrix of $\vec F(\vec x)$ defined as
\begin{equation}
\delta\vec F(\vec x)=\tens J (\vec x) \delta\vec x .
\end{equation}
Then the correction to the solution in the $k$-th iteration will be
calculated solving the linear system
\begin{equation}
\tens J(\vec x^{(k)})\delta \vec x^{(k)}=-\delta \vec F(\vec x^{(k)})
\end{equation}
and we set $\vec x^{(k+1)}=\vec x^{(k)}+\delta \vec x^{(k)}$.

With an appropriate initial approximation $\vec x^{(0)}$,
Newton's method has quadratic convergence. In practice, a rotating stellar model
can be calculated in approximately 10 iterations starting with the
corresponding non-rotating model.

\subsubsection{The case of flow fields}

The computation of the flow field is certainly the most delicate part
of the solution. Its computation needs to circumvent two difficulties:
first the flow faces very large variations of the density (typically
eight orders of magnitude) and second the low viscosity of the stellar
fluid. Indeed, this latter complication implies that the flows need to
be computed within the asymptotic regime of low Ekman numbers while
the zero viscosity solution is degenerate for the linear part of the
velocity operator (any geostrophic flow may be added to a solution see
\citealt{R06} for detailed explanations). We shall present this rather
technical point in a separate work \cite[][]{ELR12}.

\subsection{Tests of the results}

We checked the results with two global tests:
the virial theorem and the global balance of energy.

\subsubsection{The virial test}

Let us first present the virial test. For this, we recall that the equations
of a steady flow in a rotating frame are:
\begin{align*}
  2\vO\wedge\rho\vu + \rho\vu\cdot\na\vu &= 
  -\rho\na\phi+\rho\Omega^2s\es + {\rm\bf Div}[\sigma]\\
  \Div\rho\vu &= 0
\end{align*}
with the boundary conditions on the velocity field $\vu\cdot\vn=0$ and
$\vn\wedge[\sigma]\vn=\vzero$. The virial equality is obtained by
integrating the scalar product of $\vr$ with the momentum equation over
the fluid's volume.  Hence, we have to evaluate the following integrals:

\begin{itemize}
\item The $z$-component of the relative angular momentum
\[ 
\intvol \vr\cdot2\vO\wedge\rho\vu \,dV =
-2\vO\cdot\intvol\vr\wedge\rho\vu \,dV = -2\Omega L_z.
\]

\item The gravitational energy
\[ 
-\intvol \vr\cdot\rho\na\phi \,dV = \demi\intvol\rho\phi\,\dV = W.
\]

\item The kinetic energy due to bulk rotation as measured by the frame
rotation
\[ 
\intvol \vr\cdot \rho\Omega^2s\es \,dV = \intvol \rho s^2\,dV = I\Omega^2,
\]
where $I$ is the moment of inertia.

\item The relative kinetic energy
\[ 
\intvol \vr\cdot\rho\vu\cdot\na\vu \,dV = -\intvol\rho u^2\,dV =-2T_{\rm rel} .
\]

\item The stress integral
\[ 
\intvol \vr\cdot{\rm\bf Div}[\sigma] \,dV 
= \intsur r_i\sigma_{ij}\,dS_j - \intvol \sigma_{ii}\,dV .
\]
Stellar gas is assumed to be a newtonian fluid without volume
viscosity, hence the stress tensor is 
\[ 
\sigma_{ij}=\mu c_{ij}-p\delta_{ij}
\]
where $[c]$ is the shear tensor
($c_{ij}=\partial_iv_j+\partial_jv_i-2(\partial_kv_k)\delta_{ij}/3$),
so that $\sigma_{ii}=-3p$. 

\end{itemize}
Thus the virial equality may be written
\[ 
2T_{\rm rel}+I\Omega_*^2 +W + 3P + I_{s} + 2\Omega_* L_z = 0
\]
or, with non-dimensional quantities
\beq 
2T_{\rm rel}+I\Omega^2 +\Lambda_pW + 3P + I_{s} + 2\Omega L_z = 0
\eeq
where $I_s$ is the surface integral that appears in the stress
integral. In the case of a steady configuration like the one under
consideration, the surface integral is estimated as 
follows:
\[ 
I_s = \intsur \mu r_ic_{ij}\,dS_j - \intsur p\vr\cdot\dS 
\]
with
\[ 
\vn=\vE^\zeta/\left\|\vE^\zeta\right\|, \qquad
\vr\cdot\vn=\frac{r}{r_\zeta\sqrt{g^{\zeta\zeta}}}, \qquad 
dS = r\sqrt{r^2+r_\theta^2}\,d\Omega .
\]
Using nondimensional quantities, this leads to
\[ 
I_s = \isph\lc E\lp 2\partial_ru_r-\frac{2}{3}\Div\vu-
\frac{r_\theta}{r}\lp\drtheta{u_r}+\dr{u_\theta}-\frac{u_\theta}{r}\rp\rp
- p(\theta)\rc r^3\,d\Omega .
\]

\subsubsection{The energy test}

Another test of internal coherence of the solutions is provided by the
energy balance between sources and losses. The integral of the entropy 
equation over the fluid's volume gives:
\[
\intvol \rho T\vu\cdot\na s \,dV = 
\intsur (\khi\na T+\khi_tT\na s)\cdot\dS + \intvol \eps \,dV . 
\]
In the case of radiative envelopes at zero Prandtl number, this equation
can be simplified to 
\[  
\intsur (\khi\na T)\cdot\dS + \intvol \eps \,dV = 0 .
\]

\section{Some results}
The foregoing algorithm has been used to compute models of rotating stars
in a steady state. For these models we use an analytic expression for
the energy generation rate, namely:
\[ 
\eps_*(\rho,T,X,Z) = \eps_0(X,Z)\rho^2T^{-2/3}\exp\lp A/T^{1/3}\rp 
\]
as in \cite{ELR07}. It is completed by the use of OPAL tables for
the computation of the opacity and the derivation of the density 
from the equation of state ($X=0.7$ and $Z=0.02$ with solar 
composition of \citealt{GN93}).

We computed a few models of stars with a convective core, assumed to
be an isentropic region, and with a radiative envelope.  Presently, no
convective envelope can be included in the models, which are therefore
limited to stars with masses above 1.5~M$_\odot$.

\begin{figure}[t]
  \includegraphics[width=\textwidth]{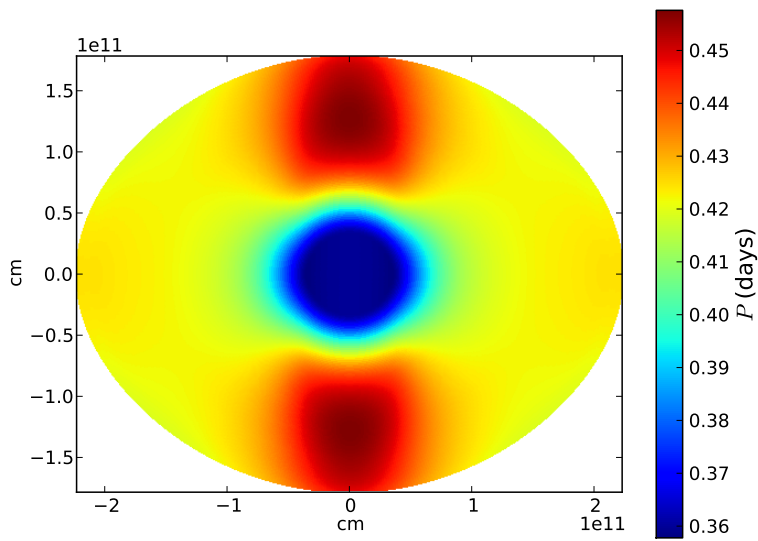}
  \caption{Differential rotation in a 5 M$_\odot$ stellar model rotating
    at $\Omega=0.7\Omega_K$.\label{rot_diff}}
\end{figure}

\begin{figure}[t]
  \begin{minipage}[t]{0.48\linewidth}
    \includegraphics[width=\textwidth]{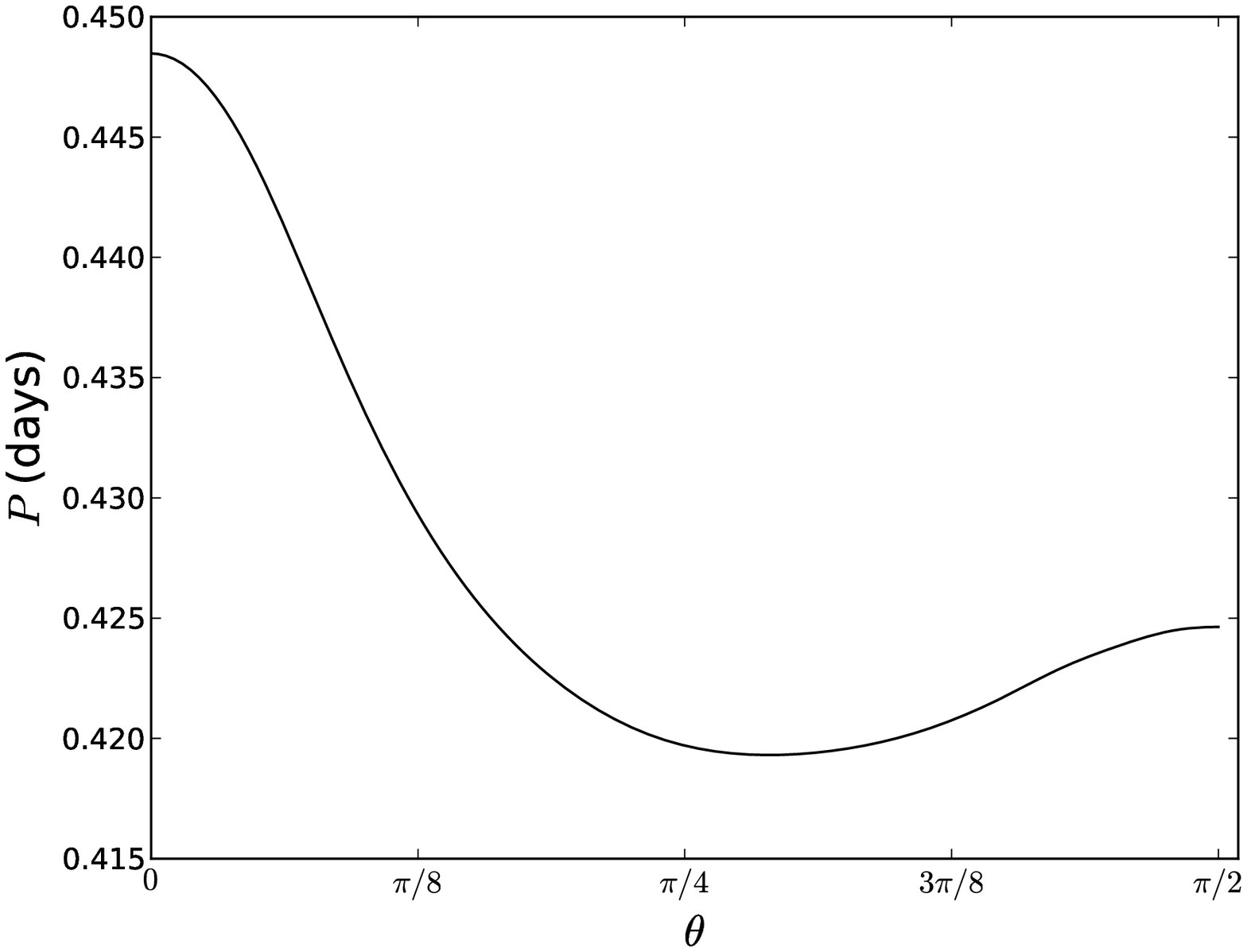}
    \caption{The surface rotation as a function of colatitude. The stellar
      model is the same as that of Fig.~\ref{rot_diff}.\label{fig_rs}}
  \end{minipage}\hfill
  \begin{minipage}[t]{0.50\linewidth}
    \includegraphics[width=\textwidth]{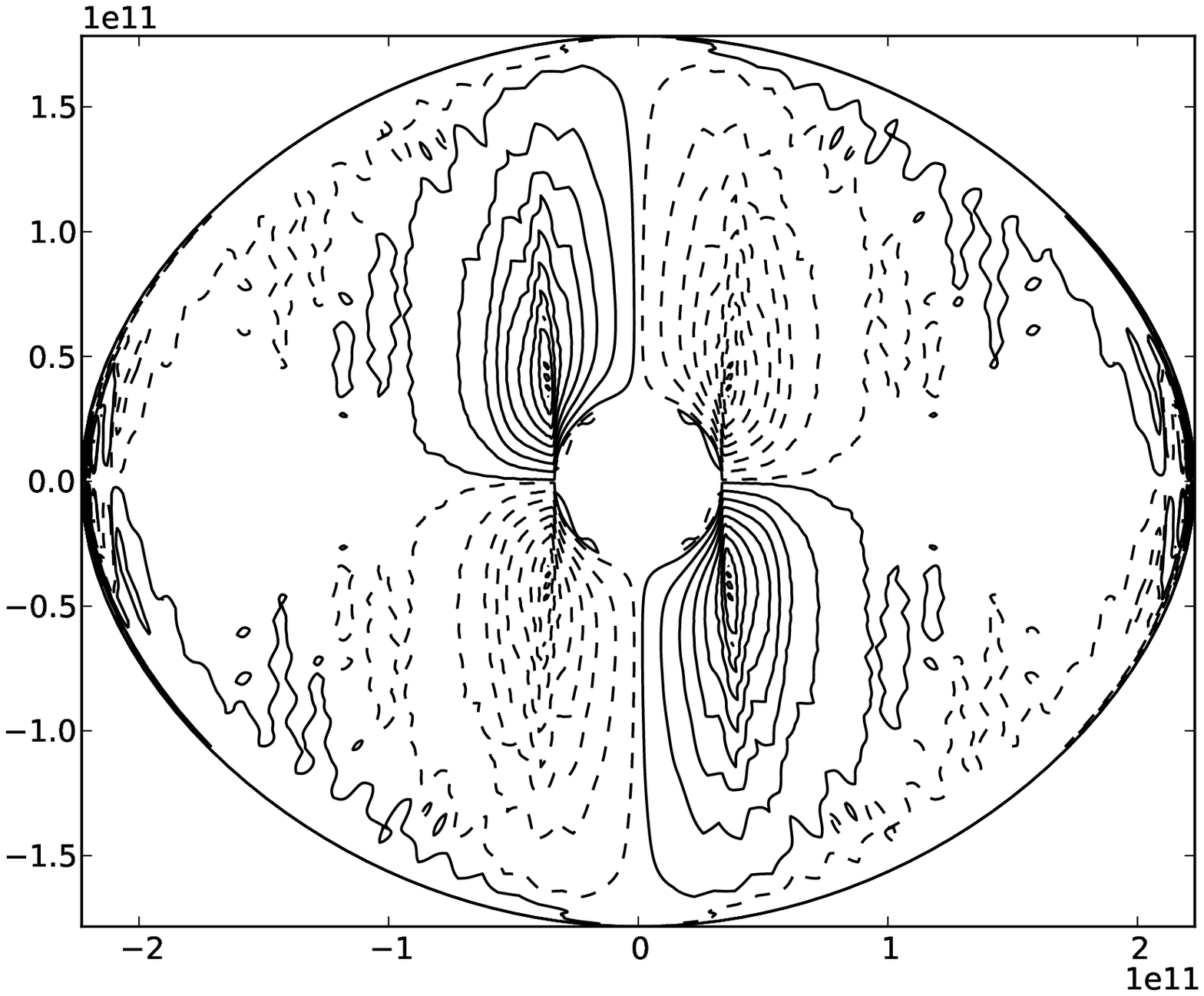}
    \caption{Streamlines of the meridian circulation associated with the
      differential rotation shown in Fig.~\ref{rot_diff}. Solid lines show
      counterclockwise circulation and dashed lines clockwise one.\label{circ}}
  \end{minipage}
\end{figure}

Numerical difficulties come both from the high density and pressure
contrast between centre and from high rotation rates. At the 
time of writing (April 2012) the most extreme models deal with
\[ 
p_{\rm surf}/p_{\rm centre} = 10^{-14} \andet \Omega/\Omega_K = 0.9 .
\]
The latter rotation corresponds to a flattening of 0.3. For these
models, the spatial resolution uses a spherical harmonics series
truncated at $L=64$ and $N=400$ collocation points on the radial 
grid, which are distributed over 8 domains.

As far as the velocity field is concerned, we recall that the baroclinic
flows that pervade the radiative region are computed in the asymptotic
limit of vanishing Ekman numbers. Thus, the Prandtl number is also set
to zero and heat advection by meridional flows is neglected.
Fig.~\ref{rot_diff} illustrates the differential rotation that is forced
by the baroclinic torque. As also observed in previous (simpler) models
of \cite{ELR07} and \cite{REL09}, the core is rotating faster
than the envelope. The differential rotation is cylindrical in the
isentropic core (as required by the Taylor-Proudman theorem), and almost
shellular in the inner part of the radiative envelope.

The meridional circulation shown in Fig.~\ref{circ} is dominated by the
streamlines of the Stewartson layers lying along the tangent cylinder of
the convective core. This flow pattern should not be taken at face value
since the interior flows are computed without viscosity. The balance of
forces in the Stewartson layer cannot be ensured
and therefore the flow pattern depends on the grid resolution.

\begin{figure}[t]
\includegraphics[width=\textwidth]{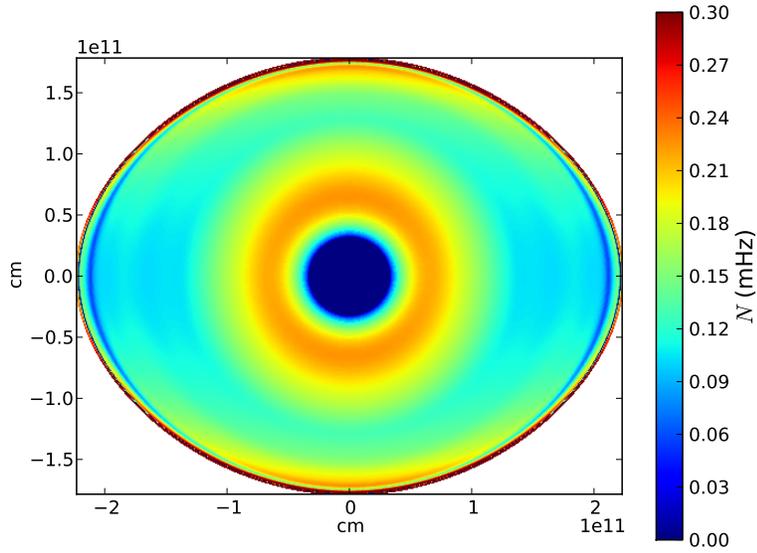}
\caption{The square of the \BVF\ distribution in a meridian plane.}
\label{fig_n2}
\end{figure}

In Fig.~\ref{fig_n2}, we provide a view of the squared \BVF. This shows
the anisotropy of the buoyancy force which, especially in the outer
layers, influences the gravito-inertial modes that are part of the
oscillation spectrum of such stars \cite[][]{DR00}.

\begin{figure}[t]
\includegraphics[width=\textwidth]{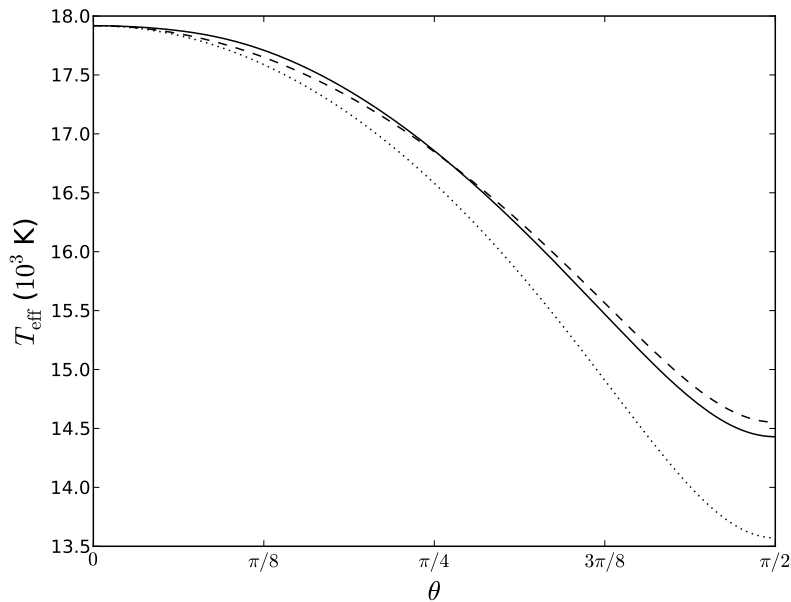}
\caption{Latitudinal flux variation (gravity darkening) at the surface
of a rapidly rotating stellar model of 5\SM. The solid line represents the
full two-dimensional model, the dashed line the simplest model described
in text and in \cite[][]{ELR11}, while the dotted line shows the
prediction of von Zeipel hypothesis.\label{flux}}
\end{figure}

Besides the dynamics, we have also investigated the thermal structure of
rotating stars and, more specifically, the distribution of the heat flux
as a function of latitude. In \cite{ELR11}, 2D models have been used to
validate a very simple model of the latitudinal variations of the flux,
which depends on a single parameter $\Omega/\Omega_K$. Such models are
based on the idea that within an envelope the heat flux $\vF$ satisfies
$\Div\vF=0$ and is almost anti-parallel to the local effective gravity
$\vg_e$. Since the mass distribution inside massive stars is concentrated
the Roche model can be used, leading to latitudinal flux variations
that only depend on the rotation rate.  This simple model has been
successfully compared with the very few observations that are available
and with complete two-dimensional models \cite[see][]{ELR11}. As shown in
Fig.~\ref{flux} the two models nicely match and notably differ from the
model of von Zeipel which predicts that T$_{\rm eff}\propto g_e^{1/4}$
as a consequence of neglecting of the baroclinicity of the configuration.

In Tables~\ref{model_grid} and \ref{model_grid2} we show the physical
parameters obtained from the calculations of a series of stellar models
with masses between 3 and 20 $M_\odot$ and rotation rates up to 90\%
of the breakup angular velocity.  The models have been calculated
using $L_\mathrm{max}=64$ and 400 radial points distributed over 8
domains. The virial and energy tests give an idea of the quality of the
solutions. As we can see, the virial test values are very small, below a
few $10^{-10}$, as a result of the high precision of spectral methods. The
energy test is also good enough (under around
$10^{-5}$), the larger values being a consequence of the use of tabulated
opacities which have limited precision.

\section{Conclusions}

The computation of two-dimensional stellar models is a real numerical
challenge which has been taken  up by the ESTER project. 
The key difficulty is to derive simultaneously the bulk structure of the star 
and the mean velocity fields that pervade it.
These large-scale flows
come from either the baroclinicity of the radiative zones or from the
anisotropic Reynolds stresses in the convective zones (although some
Reynolds stresses might also be expected in the radiative zones if shear
instabilities can generate some small-scale turbulence as suggested by
\citealt{zahn92}). At the moment, the ESTER code can produce dynamically
self-consistent models, which include the background flows but no Reynolds
stresses, for stars with masses larger than 1.5\SM\ and rotation rates
less than 90\% the break-up. The possibilities of the code have not yet been
fully explored however.


\input{table}

\input{table2}

As far as steady solutions are concerned, the main challenges are to
enable the modelling of the lower-mass stars (solar type) with an outer
convection zone, and to take into account the effects of viscosity and
Reynolds stresses in the bulk of the stars. These latter effects might
indeed be crucial to the transport of elements.

The next important step will be to deal with time evolution; this
should include:
\begin{itemize}
\item dynamical evolution during PMS phase with a slow gravitational
contraction inducing spin-up of the star;
\item dynamical evolution during main sequence with stellar wind, which
forces a spin down and associated mixing;
\item nuclear evolution.
\end{itemize}

These steps need the right algorithm for temporal evolution, which is not
known presently. Indeed, such an algorithm should be at the same time
fast, stable and precise. We have managed to use spectral methods, which
ensure rapidity and precision but the stability 
remains a challenge. A better understanding of the properties of the 
discretized operators is certainly a key to improve the efficiency of 
the algorithms.

As shown by the foregoing examples, some realistic models can now be
computed for intermediate mass and massive stars. These models are steady
and therefore the chemical composition must be given. To circumvent
this constraint, one-dimensional models can be used to compute time
evolution and the ensuing chemical composition. Then, the bulk relation
between pressure and chemical composition shown by the 1D-model can be
inserted into the 2D-model. In this way, steady models can include some
stellar evolution.

The steady models described in this work are most relevant in the study
of the oscillation properties of rotating stars. The interpretation of
the frequency spectrum of such stars is indeed a challenging problem and
an intense use of 2D-models will be necessary to find out how to invert
data coming from, for instance, $\delta$-Scuti stars.

Another use of these models is obviously the interpretation of
interferometric data collected on some nearby fast rotating stars
($\alpha$~Aql, $\alpha$~Cep, $\alpha$~Leo, $\beta$~Cas, etc.). Fast
rotating stars have a surface brightness that strongly depends on
latitude (gravity darkening). Accurate models are crucial to extract
the physical parameters contained in the interferometric data from 
such stars.

Finally, steady two-dimensional models may also serve as proxies for the
initial conditions of a collapsing massive star, although the final word
will come from time-evolved models including mass-loss.

\begin{acknowledgement}
The authors acknowledge the support of the French Agence Nationale de
la Recherche (ANR), under grant ESTER (ANR-09-BLAN-0140).  This work
was also supported by the Centre National de la Recherche Scientifique
(CNRS, UMR 5277), through the Programme National de Physique Stellaire
(PNPS). The numerical calculations were carried out on the CalMip
machine of the ``Centre Interuniversitaire de Calcul de Toulouse''
(CICT) which is gratefully acknowledged.
\end{acknowledgement}

\bibliographystyle{aa}
\bibliography{/home/virgo/tex/biblio/bibnew.bib}

\end{document}

%% file: table.tex
\begin{table}
\caption{Fundamental parameters for a series of rotating stellar models$^a$.}
\label{model_grid}
\begin{tabular}{ p{0.9cm} p{0.9cm} p{1.1cm} p{0.7cm} p{1.1cm} p{1.1cm} p{1.2cm} p{1.2cm} p{1.4cm} p{1cm} }
\hline\noalign{\smallskip}
$\!\!$M(M$_\odot$)  &
$\Omega/\Omega_K^b$  &
R(R$_\odot$)  &
$\varepsilon$ $^c$  &
$P_\mathrm{rot}^\mathrm{s}$(d)  &
$P_\mathrm{rot}^\mathrm{c}$(d)  &
$\!\!v_\mathrm{eq}${\scriptsize (km/s)}  &
$L(L_\odot)$  &
$\!\!T_\mathrm{eff}$($10^3$K)  &
$\log g_e$  \\
\noalign{\smallskip}\svhline\noalign{\smallskip}
3.0 & 0.0 & 1.97 & 0.00 &  - &  - & 0.0 & 81.2 & 12.36 & 4.33 \\ \noalign{\smallskip}
3.0 & 0.3 & 1.96(p)\newline 2.05(e)& 0.04 & 0.70(p)\newline 0.65(e)& 0.54 & 158.6 & 80.0 & 12.50(p)\newline 11.97(e) & 4.33(p)\newline 4.25(e)\\
\noalign{\smallskip}
3.0 & 0.5 & 1.95(p)\newline 2.19(e)& 0.11 & 0.46(p)\newline 0.43(e)& 0.36 & 255.5 & 78.4 & 12.69(p)\newline 11.31(e) & 4.34(p)\newline 4.11(e)\\
\noalign{\smallskip}
3.0 & 0.7 & 1.94(p)\newline 2.42(e)& 0.20 & 0.38(p)\newline 0.36(e)& 0.30 & 340.4 & 77.0 & 12.84(p)\newline 10.36(e) & 4.34(p)\newline 3.86(e)\\
\noalign{\smallskip}
3.0 & 0.9 & 1.93(p)\newline 2.74(e)& 0.29 & 0.35(p)\newline 0.34(e)& 0.28 & 411.5 & 76.4 & 12.92(p)\newline 8.91(e) & 4.34(p)\newline 3.32(e)\\
\noalign{\smallskip}
\hline\noalign{\smallskip}
5.0 & 0.0 & 2.62 & 0.00 &  - &  - & 0.0 & 542.8 & 17.23 & 4.30 \\ \noalign{\smallskip}
5.0 & 0.3 & 2.60(p)\newline 2.72(e)& 0.04 & 0.83(p)\newline 0.77(e)& 0.66 & 177.7 & 533.2 & 17.44(p)\newline 16.69(e) & 4.31(p)\newline 4.23(e)\\
\noalign{\smallskip}
5.0 & 0.5 & 2.58(p)\newline 2.91(e)& 0.11 & 0.55(p)\newline 0.51(e)& 0.44 & 286.5 & 520.9 & 17.70(p)\newline 15.76(e) & 4.31(p)\newline 4.09(e)\\
\noalign{\smallskip}
5.0 & 0.7 & 2.57(p)\newline 3.21(e)& 0.20 & 0.45(p)\newline 0.42(e)& 0.36 & 381.9 & 510.4 & 17.92(p)\newline 14.43(e) & 4.32(p)\newline 3.83(e)\\
\noalign{\smallskip}
5.0 & 0.9 & 2.56(p)\newline 3.63(e)& 0.30 & 0.42(p)\newline 0.40(e)& 0.33 & 461.6 & 505.3 & 18.02(p)\newline 12.40(e) & 4.32(p)\newline 3.30(e)\\
\noalign{\smallskip}
\hline\noalign{\smallskip}
10.0 & 0.0 & 3.87 & 0.00 &  - &  - & 0.0 & 5733.6 & 25.55 & 4.26 \\ \noalign{\smallskip}
10.0 & 0.3 & 3.84(p)\newline 4.02(e)& 0.04 & 1.07(p)\newline 0.98(e)& 0.86 & 206.8 & 5619.0 & 25.85(p)\newline 24.73(e) & 4.27(p)\newline 4.19(e)\\
\noalign{\smallskip}
10.0 & 0.5 & 3.81(p)\newline 4.29(e)& 0.11 & 0.70(p)\newline 0.65(e)& 0.57 & 333.6 & 5469.6 & 26.24(p)\newline 23.35(e) & 4.27(p)\newline 4.05(e)\\
\noalign{\smallskip}
10.0 & 0.7 & 3.78(p)\newline 4.73(e)& 0.20 & 0.57(p)\newline 0.54(e)& 0.46 & 444.8 & 5341.5 & 26.57(p)\newline 21.35(e) & 4.28(p)\newline 3.80(e)\\
\noalign{\smallskip}
10.0 & 0.9 & 3.76(p)\newline 5.37(e)& 0.30 & 0.53(p)\newline 0.51(e)& 0.43 & 536.7 & 5279.3 & 26.73(p)\newline 18.28(e) & 4.28(p)\newline 3.26(e)\\
\noalign{\smallskip}
\hline\noalign{\smallskip}
20.0 & 0.0 & 5.70 & 0.00 &  - &  - & 0.0 & 43791.2 & 35.00 & 4.23 \\ \noalign{\smallskip}
20.0 & 0.3 & 5.66(p)\newline 5.91(e)& 0.04 & 1.36(p)\newline 1.24(e)& 1.13 & 241.0 & 42921.1 & 35.41(p)\newline 33.89(e) & 4.23(p)\newline 4.16(e)\\
\noalign{\smallskip}
20.0 & 0.5 & 5.61(p)\newline 6.32(e)& 0.11 & 0.89(p)\newline 0.82(e)& 0.74 & 388.5 & 41775.4 & 35.94(p)\newline 31.97(e) & 4.24(p)\newline 4.01(e)\\
\noalign{\smallskip}
20.0 & 0.7 & 5.57(p)\newline 7.00(e)& 0.20 & 0.72(p)\newline 0.68(e)& 0.61 & 516.9 & 40790.9 & 36.39(p)\newline 29.16(e) & 4.24(p)\newline 3.76(e)\\
\noalign{\smallskip}
20.0 & 0.9 & 5.54(p)\newline 8.02(e)& 0.31 & 0.67(p)\newline 0.65(e)& 0.56 & 621.0 & 40326.9 & 36.60(p)\newline 24.82(e) & 4.25(p)\newline 3.22(e)\\
\noalign{\smallskip}
\end{tabular}
$^a$ From left to right: Mass, equatorial angular velocity, radius, flattening, 
central rotation period, surface rotation period, equatorial velocity, luminosity, 
effective temperature, logarithm of effective gravity (cgs). 
The values for the solar parameters used in the table are
$M_\odot=1.9891\cdot\!10^{33}$g, $R_\odot=6.95508\cdot\!10^{10}$cm
 and $L_\odot=3.8396\cdot\!10^{33}$erg/s.

$^b$ $\Omega_K=\sqrt{\frac{GM}{R_e^3}}\,$.

$^c$ Flattening $\varepsilon=1-\frac{R_p}{R_e}\,$.
\end{table}

%% file: table2.tex
\begin{table}
\caption{Fundamental parameters for a series of rotating stellar models (cont.)$^a$.}
\label{model_grid2}
\begin{tabular}{ p{0.9cm} p{0.8cm} p{1.65cm} p{0.8cm} p{1.3cm} p{1.5cm} p{1.5cm} p{1.5cm} p{1.2cm} }
\hline\noalign{\smallskip}
$\!\!$M(M$_\odot$)  &
$\Omega/\Omega_K$  &
$p_c$(dyn/cm$^2)$  &
$\rho_c$(g)  &
$T_c$(K)  &
$p_s/p_c$  &
$\rho_s/\rho_c$  &
Virial\newline test $^b$  &
Energy\newline test $^c$  \\
\noalign{\smallskip}\svhline\noalign{\smallskip}
3.0 & 0.0 & $1.34\cdot\! 10^{17}$ & $40.8$ & $2.43\cdot\! 10^{7}$ & $9.61\cdot\! 10^{\mbox{-}15}$ & $2.03\cdot\! 10^{\mbox{-}11}$ & $5.08\cdot\! 10^{\mbox{-}11}$ & $5.84\cdot\! 10^{\mbox{-}7}$ \\ \noalign{\smallskip}
3.0 & 0.3 & $1.34\cdot\! 10^{17}$ & $41.0$ & $2.43\cdot\! 10^{7}$ & $9.92\cdot\! 10^{\mbox{-}15}$ & $2.07\cdot\! 10^{\mbox{-}11}$ & $8.01\cdot\! 10^{\mbox{-}11}$ & $3.42\cdot\! 10^{\mbox{-}7}$ \\ \noalign{\smallskip}
3.0 & 0.5 & $1.35\cdot\! 10^{17}$ & $41.3$ & $2.42\cdot\! 10^{7}$ & $1.03\cdot\! 10^{\mbox{-}14}$ & $2.09\cdot\! 10^{\mbox{-}11}$ & $2.03\cdot\! 10^{\mbox{-}10}$ & $6.14\cdot\! 10^{\mbox{-}6}$ \\ \noalign{\smallskip}
3.0 & 0.7 & $1.35\cdot\! 10^{17}$ & $41.5$ & $2.42\cdot\! 10^{7}$ & $1.07\cdot\! 10^{\mbox{-}14}$ & $2.08\cdot\! 10^{\mbox{-}11}$ & $2.67\cdot\! 10^{\mbox{-}10}$ & $7.50\cdot\! 10^{\mbox{-}6}$ \\ \noalign{\smallskip}
3.0 & 0.9 & $1.36\cdot\! 10^{17}$ & $41.6$ & $2.42\cdot\! 10^{7}$ & $1.09\cdot\! 10^{\mbox{-}14}$ & $1.93\cdot\! 10^{\mbox{-}11}$ & $2.31\cdot\! 10^{\mbox{-}10}$ & $1.09\cdot\! 10^{\mbox{-}5}$ \\ \noalign{\smallskip}
\hline\noalign{\smallskip}
5.0 & 0.0 & $8.00\cdot\! 10^{16}$ & $21.3$ & $2.76\cdot\! 10^{7}$ & $3.86\cdot\! 10^{\mbox{-}14}$ & $6.22\cdot\! 10^{\mbox{-}11}$ & $1.20\cdot\! 10^{\mbox{-}10}$ & $3.75\cdot\! 10^{\mbox{-}6}$ \\ \noalign{\smallskip}
5.0 & 0.3 & $8.04\cdot\! 10^{16}$ & $21.4$ & $2.75\cdot\! 10^{7}$ & $3.97\cdot\! 10^{\mbox{-}14}$ & $6.30\cdot\! 10^{\mbox{-}11}$ & $8.41\cdot\! 10^{\mbox{-}11}$ & $8.84\cdot\! 10^{\mbox{-}6}$ \\ \noalign{\smallskip}
5.0 & 0.5 & $8.09\cdot\! 10^{16}$ & $21.6$ & $2.75\cdot\! 10^{7}$ & $4.11\cdot\! 10^{\mbox{-}14}$ & $6.34\cdot\! 10^{\mbox{-}11}$ & $7.51\cdot\! 10^{\mbox{-}11}$ & $7.95\cdot\! 10^{\mbox{-}6}$ \\ \noalign{\smallskip}
5.0 & 0.7 & $8.13\cdot\! 10^{16}$ & $21.7$ & $2.75\cdot\! 10^{7}$ & $4.23\cdot\! 10^{\mbox{-}14}$ & $6.21\cdot\! 10^{\mbox{-}11}$ & $8.01\cdot\! 10^{\mbox{-}11}$ & $7.33\cdot\! 10^{\mbox{-}6}$ \\ \noalign{\smallskip}
5.0 & 0.9 & $8.16\cdot\! 10^{16}$ & $21.8$ & $2.74\cdot\! 10^{7}$ & $4.29\cdot\! 10^{\mbox{-}14}$ & $5.58\cdot\! 10^{\mbox{-}11}$ & $1.06\cdot\! 10^{\mbox{-}10}$ & $1.34\cdot\! 10^{\mbox{-}5}$ \\ \noalign{\smallskip}
\hline\noalign{\smallskip}
10.0 & 0.0 & $4.31\cdot\! 10^{16}$ & $9.44$ & $3.20\cdot\! 10^{7}$ & $1.53\cdot\! 10^{\mbox{-}13}$ & $1.79\cdot\! 10^{\mbox{-}10}$ & $8.96\cdot\! 10^{\mbox{-}12}$ & $1.12\cdot\! 10^{\mbox{-}6}$ \\ \noalign{\smallskip}
10.0 & 0.3 & $4.33\cdot\! 10^{16}$ & $9.50$ & $3.20\cdot\! 10^{7}$ & $1.56\cdot\! 10^{\mbox{-}13}$ & $1.80\cdot\! 10^{\mbox{-}10}$ & $4.91\cdot\! 10^{\mbox{-}11}$ & $5.53\cdot\! 10^{\mbox{-}7}$ \\ \noalign{\smallskip}
10.0 & 0.5 & $4.35\cdot\! 10^{16}$ & $9.58$ & $3.19\cdot\! 10^{7}$ & $1.60\cdot\! 10^{\mbox{-}13}$ & $1.78\cdot\! 10^{\mbox{-}10}$ & $5.18\cdot\! 10^{\mbox{-}11}$ & $1.25\cdot\! 10^{\mbox{-}6}$ \\ \noalign{\smallskip}
10.0 & 0.7 & $4.38\cdot\! 10^{16}$ & $9.65$ & $3.19\cdot\! 10^{7}$ & $1.64\cdot\! 10^{\mbox{-}13}$ & $1.69\cdot\! 10^{\mbox{-}10}$ & $2.59\cdot\! 10^{\mbox{-}11}$ & $5.62\cdot\! 10^{\mbox{-}7}$ \\ \noalign{\smallskip}
10.0 & 0.9 & $4.39\cdot\! 10^{16}$ & $9.69$ & $3.18\cdot\! 10^{7}$ & $1.66\cdot\! 10^{\mbox{-}13}$ & $1.34\cdot\! 10^{\mbox{-}10}$ & $4.54\cdot\! 10^{\mbox{-}11}$ & $4.48\cdot\! 10^{\mbox{-}7}$ \\ \noalign{\smallskip}
\hline\noalign{\smallskip}
20.0 & 0.0 & $2.79\cdot\! 10^{16}$ & $4.88$ & $3.62\cdot\! 10^{7}$ & $3.26\cdot\! 10^{\mbox{-}13}$ & $2.41\cdot\! 10^{\mbox{-}10}$ & $1.54\cdot\! 10^{\mbox{-}11}$ & $2.38\cdot\! 10^{\mbox{-}6}$ \\ \noalign{\smallskip}
20.0 & 0.3 & $2.80\cdot\! 10^{16}$ & $4.91$ & $3.61\cdot\! 10^{7}$ & $3.31\cdot\! 10^{\mbox{-}13}$ & $2.36\cdot\! 10^{\mbox{-}10}$ & $7.25\cdot\! 10^{\mbox{-}11}$ & $1.49\cdot\! 10^{\mbox{-}6}$ \\ \noalign{\smallskip}
20.0 & 0.5 & $2.81\cdot\! 10^{16}$ & $4.94$ & $3.61\cdot\! 10^{7}$ & $3.39\cdot\! 10^{\mbox{-}13}$ & $2.23\cdot\! 10^{\mbox{-}10}$ & $1.79\cdot\! 10^{\mbox{-}11}$ & $2.11\cdot\! 10^{\mbox{-}7}$ \\ \noalign{\smallskip}
20.0 & 0.7 & $2.82\cdot\! 10^{16}$ & $4.98$ & $3.60\cdot\! 10^{7}$ & $3.46\cdot\! 10^{\mbox{-}13}$ & $1.93\cdot\! 10^{\mbox{-}10}$ & $7.16\cdot\! 10^{\mbox{-}11}$ & $4.38\cdot\! 10^{\mbox{-}8}$ \\ \noalign{\smallskip}
20.0 & 0.9 & $2.83\cdot\! 10^{16}$ & $5.00$ & $3.60\cdot\! 10^{7}$ & $3.49\cdot\! 10^{\mbox{-}13}$ & $1.17\cdot\! 10^{\mbox{-}10}$ & $4.09\cdot\! 10^{\mbox{-}11}$ & $1.09\cdot\! 10^{\mbox{-}6}$ \\ \noalign{\smallskip}
\end{tabular}
$^a$ From left to right: Mass, equatorial angular velocity, central pressure, 
central density, central temperature, ratio between surface and central pressure, 
ratio between surface and central density, relative virial test and relative energy test. 
The values for the solar parameters used in the table are
$M_\odot=1.9891\cdot\!10^{33}$g, $R_\odot=6.95508\cdot\!10^{10}$cm
 and $L_\odot=3.8396\cdot\!10^{33}$erg/s.

$^b$ Virial test: $(2T_\mathrm{rel}+I\Omega_\star^2+W+3P+I_s+2\Omega_\star L_z)/W$.

$^c$ Energy test: $\left(\int_{(S)}(\xi\nabla T)\cdot \D\vec S + 
\int_{(V)}\varepsilon\D V \right)/\int_{(V)}\varepsilon\D V $
\end{table}